# Toward an Ion-Based Large-Scale Integrated Circuit:

# Circuit Level Design, Simulation, and Integration of Iontronic Components

Noa Edri Fraiman∗ , Barak Sabbagh†‡, Gilad Yossifon ‡ and Alexander Fish*

Email: noa.edri@biu.ac.il baraksabbagh@campus.technion.ac.il gyossifon@tauex.tau.ac.il alexander.fish@biu.ac.il

∗ Faculty of Engineering, Bar-Ilan University, Ramat Gan 5290002, Israel

†Faculty of Mechanical Engineering,Technion−Israel Institute of Technology, Haifa 3200003,Israel

‡ School of Mechanical Engineering, Tel Aviv University, Tel Aviv 69978, Israel

## Abstract

Iontronics combines ions as charge carriers with electronic-like operations, enabling unique information processing, chemical regulation, and enhanced bio-integrability. Standard simulation tools encounter difficulties in effectively modeling the behavior of integrated iontronic components, highlighting the need for specialized design and simulation approaches. This paper presents a design methodology for iontronic integrated circuits, inspired by well-established electronic design methodologies and made possible by the development of a compact model for the iontronic bipolar diode. Grounded in the diode's physical properties and observed behavior, this model provides a conceptual framework that could be applied to other iontronic components. It is implemented using standard VLSI (Very Large-Scale Integration) electronic design tools, enabling simulations that demonstrate diode-based iontronic circuit behaviors and laying the groundwork for the design and simulation of hybrid systems integrating electronic and iontronic circuits. The proposed iontronic circuit simulation approach enables the exploration of how component uniformity influences circuit behavior, as well as the impact of diode parameters and a deeper understanding of diode characteristics from a circuit perspective. These insights are expected to contribute to the development of more complex and efficient iontronic circuits, bringing us closer to practical and groundbreaking applications in the field.

## Introduction

The ability to integrate a large number of electronic components onto a single small piece of semiconductor, and the invention of Integrated Circuit (IC) have changed the world beyond imagination. The lengthy trajectory from the invention of the bi-polar transistor to today's integration capabilities has prompted the development of new fields specializing in new devices [1], [2], [3], advanced fabrication processes for integration [4], simulation tools and Electronic Design Automation (EDA) to allow for reliable chip design [5], [6], and ongoing miniaturization to advance integrated circuit capabilities [7], [8], [9].

In recent years, new and exciting computational domains have been proposed and developed, with the potential to expand solid-state electronics [10], [11], [12]. Iontronics, an emerging cutting-edge field, explores the ability to perform computations using ions that serve as information carriers, and is based on the elegant functionalities found in biological systems [13]. This novel approach leverages the unique ability of nanoscale fluidic structures such as nanochannels and ion exchange membranes to selectively transport ions based on their charge. In fluidic nanostructures, the overlapping electric double layers (EDLs) cause co-ions (related to the nanostructure's surface charge) to be excluded from the inner volume, while counter-ions can transport freely. As a result, modulating the surface charge or functionalized molecular probes enables control over the flux of specific ions and molecules when an external electric field is applied [14], [15]. Combining various micro- and nanofluidic structures that exhibit different ion perm-selective behaviors, ranging from low to high perm-selectivity to either anions or cations, enables the creation of novel ion-based computing elements such as iontronic resistors, diodes [16], and transistors [17] that have ushered in a new era of information processing.

The interest in exploring the integration capability of iontronic components on a single chip stem from the possibilities it provides to perform more complex computations and develop alternative processing capabilities. Greater computational complexity could augment the capacities of numerous applications. These include smart drug [18] and ion [19] delivery and complex sensors arrays that require sensing and parallel processing in real time without necessarily needing to convert the information to the electronic domain [16]. In addition, since ions serve as the charge carriers in iontronics devices, and there are different types of ions in nature, iontronic circuits can provide a "colorful" signal [20] [13]. Therefore, computational elements built from iontronic devices have the potential to deliver more information; for example, detecting the type of ions at the circuit's output can provide additional information about the inputs of the circuit [21] , [22]. In addition to these advantages, iontronics has a natural interface to electronics. Integrating iontronics and electronics opens up new and exciting opportunities for hybrid systems that can leverage both electrical signals based on electrons and ion-based signals transmitted through fluidic elements [23]. This synergy creates the potential for innovative applications that combine the strengths of both technologies [24].

Recent studies have demonstrated that iontronic diodes can perform Boolean computations by implementing basic logic gates such as OR and AND [16], [25], [26], [27]. Despite these advances, most studies have encountered limitations, with no more than a few logic gates integrated thus far. To the best of our knowledge, our recent study [28] , which presents an integration of five logic gates, represents the largest iontronic integrated circuit fabricated to date. Using terminology from the microelectronics world we successfully demonstrated the basic performance and challenges of these gates. The main hurdles include the design and fabrication of suitable devices with small parameter variations and high reliability, the integration and fluidic connectivity of these devices onto a single microfluidic chip, and no less importantly, a set of tools to allow efficient design and simulation of integrated computational systems to enable prediction and optimization of the complex iontronics systems already at the design level. These capabilities have existed in integrated electronic circuits since the mid-1960s [29], when scientists and engineers managed to connect dozens or hundreds of electronic devices in an integrated circuit, and reached a peak in 2024 when one of the world's most complex electronic chips comprising trillions of transistors was developed [30].

In electronic integrated circuits, the hurdle of developing complex systems can be solved by a design approach based on the creation of abstraction levels, as depicted in simplified form in Figure 1(a). Each level of abstraction has a design methodology that corresponds to that level and design, and simulation tools that support this methodology. The transition between levels of abstraction is supported by dedicated models that only describe the phenomena and characteristics that are important to the targeted level of abstraction. Technology CAD (TCAD) tools use Poisson, transport, and continuity equations to model electron and hole concentrations and current flow in semiconductor devices under specific bias conditions. However, TCAD is too complex for circuit simulation, since solving these equations for each component is highly time-consuming. For circuit-level designs, a compact model is utilized to characterize the current and charge behavior of a semiconductor device as a function of voltage, process parameters, temperature, and geometry. This model, which represents the device relative to its terminals, is incorporated into circuit simulation tools to enable both accurate and efficient simulations. [31] [29]. The transition from the circuit level to the functional and logic design adheres to a similar approach. It involves shifting the focus from detailed components and interconnects to abstract logical functions and their interactions within the system. The logic design is implemented with tools developed to handle logic gates and utilizes a gate model that only describes the significant properties of the gate such as speed, power consumption, area, etc. Although gate characteristics depend heavily on the physical properties of the device, gate models do not include these properties at all. This enables an efficient, fast, yet reliable simulation of a large number of integrated gates. In a similar way, the system-level design focuses on abstraction to manage system-wide interactions and architecture [32].

In iontronics, employing standard device simulation approaches is impractical when attempting to evaluate iontronic circuits. For example, a physically- based simulator such as COMSOL Multiphysics ® can evaluate the influence of ionic transport processes in the bipolar nanostructure that forms an iontronic diode but would face an uphill battle when trying to simulate two integrated diodes (especially if dynamic characteristics are needed). This emphasizes the urgent need for a simulation tool that can simulate iontronics-based circuits and predict their response.

Inspired by the electronic integrated circuit abstraction levels and design approach, we propose a compact model for nanofluidic diodes to initiate iontronic circuit design capabilities and lay the groundwork for circuit design methodologies for iontronic integrated circuits. The model is based on the physical characteristics of the device and its experimental behavior. It is implemented using a Verilog-A and Spectre simulator, thus enabling seamless integration and straightforward use within commercial very large-scale integration (VLSI) EDA environments. It not only predicts the behavior of iontronic circuits but can also serve as a tool for further exploration and improvement of nanofluidic devices toward larger scale integration. The development of iontronic circuit-level simulation capabilities can also facilitate the future development of iontronic gate-level models, by enabling higher abstraction tools for efficient design of large-scale iontronic integrated circuits. This model also paves the way for the simulation and integration of hybrid electronics and iontronics systems.

As depicted in Figure 1B compact models of metal-oxide-semiconductor field-effect transistor (MOSFET) transistors (the basic components of all electronic IC) start with a "core model", which only includes the basic effects [33]. However, to produce an accurate compact model, numerous add-on "real-device models" must be added. Advances in transistor fabrication technology made real-device effects dominant, making these models crucial for accurate circuit simulation [31]. For the first iontronic diode compact model we suggest a similar approach.

While we focus here on demonstrating our approach to iontronic circuit design using iontronic bipolar diodes, the method is generic and can be extended to other combinations of iontronic components, such as transistors [34], memristors [35] , resistors and capacitors [36]. In this work, we first present an iontronic diode core model that accounts for the basic effects. The model has an I-V static feature, which is crucial to calculating the steady state response of the iontronic logic gate, and a C-V feature, which describes the relationship between the voltage and the large signal capacitance to predict the switching time of the diode. Next, a very preliminary statistical model is presented as an add-on to account for random variations that take place during circuit fabrication, which significantly impact device functionality. Clearly other more advanced add-ons must be included in future models, to deal with the influence of current saturation, aging, temperature, etc. In addition, iontronics has its own unique set of effects compared to electronics, such as fluid convection and "colorful" currents; namely, currents formed by the movement of different ion types which would be valuable in future models to enhance accuracy and explore innovative iontronic circuits applications.

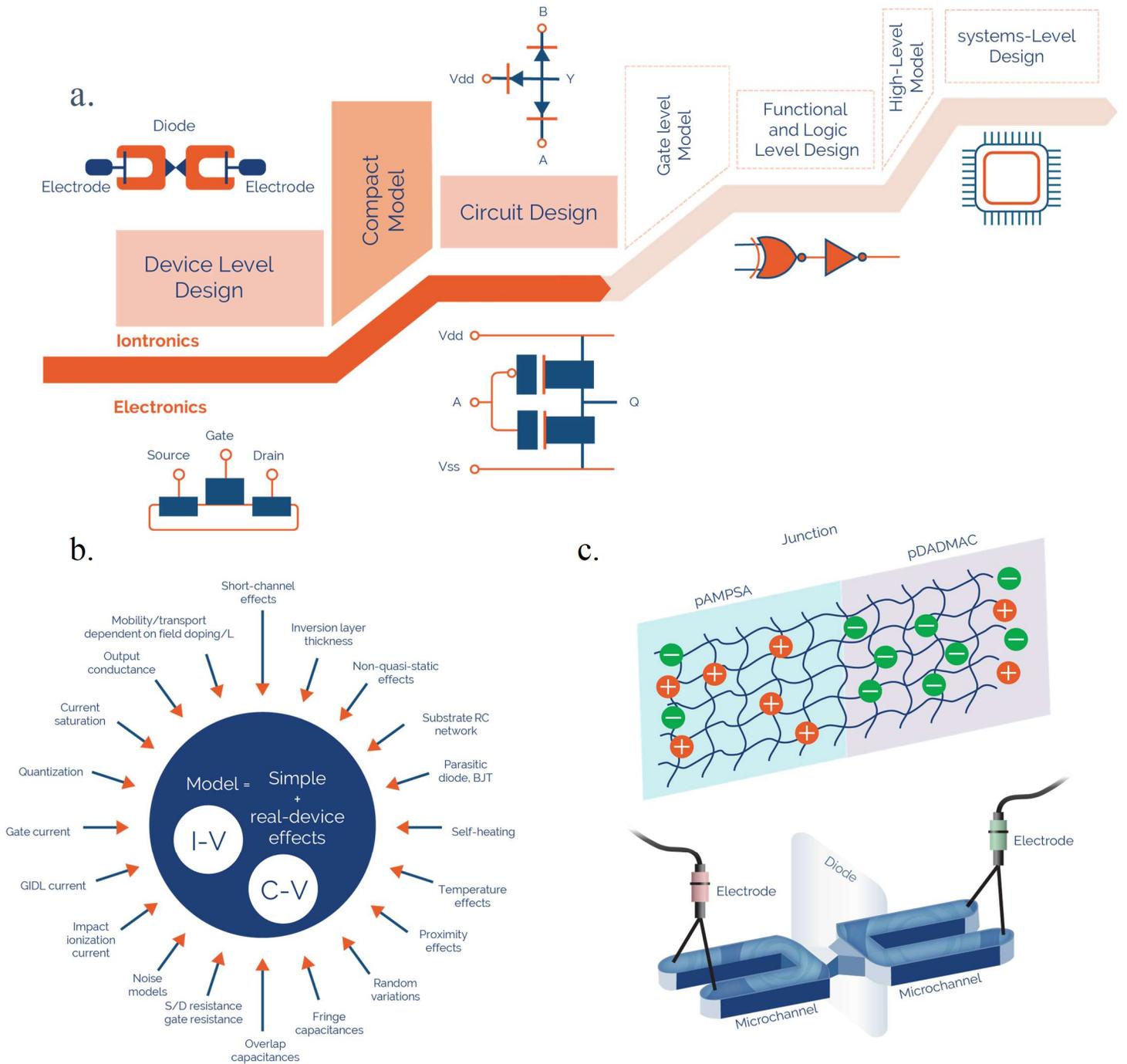

Figure 1: (a) **Abstraction Levels in the Integrated Circuit Design Process.** The bottom section of the diagram shows the conventional abstraction levels in electronic design, that move from the device level, circuit design, logic design, and system- level design. The top section suggests how to apply similar abstraction levels to iontronic systems. A and B are the inputs, Y and Q are the outputs, and Vdd and Vss are the positive and negative supply voltages, respectively. This paper focuses on the foundational step in this process, the creation of a compact model to enable iontronic circuit design.(b) **Compact Modeling Approach for Electronic Devices.** The electronic MOSFET transistor compact model approach is shown (reproduced from [33]), combining a simple model with numerous real-device models that ensure high accuracy. A similar approach is proposed for iontronic diodes as an example of iontronic components. (c) **The basic structure of the iontronic diode**, where the diode consists of interfacing cation-exchange (pAMPSA) and an anion-exchange (pDADMAC) polyelectrolytes, connected by microchannels to the electrode [28].

## Results

The development of a compact model for the iontronic diode (Figure 1(c)) was carried out in two phases.

### Phase I:

Iontronic and PN (P-type and N-type) semiconductor diodes have several similarities in structure and in the underlying mechanisms that induce asymmetric transport of oppositely charged mobile charge carriers. In both iontronic and electronic diodes, the main transport mechanisms are drift (electro-migration) and diffusion [37]. We thus used the SPICE semiconductor diode model [38], which is widely implemented in electronic integrated circuit simulators, as an inspiration for the iontronic diode compact model. The equivalent circuit for the SPICE diode model is shown in Figure 2(a). The branch current equation (BCE) describes the diode's behavior in direct current (DC) conditions and is characterized by an exponential relationship between the diode current and voltage, represented as $D_1$ in the circuit. In addition, an ohmic resistance ($R_S$) is connected in series with $D_1$. To capture the diode's dynamic behavior, the model also incorporates a diffusion capacitor ($C_d$) and a junction capacitor ($C_j$). The junction capacitance ($C_j$) is voltage-dependent because it represents the depletion region's capacitance, which varies with the applied reverse bias voltage. The diffusion capacitance ($C_d$), also known as the transit time capacitance ($C_T$), is related to the charge stored in the diode when it is forward- biased [38] .

Given the similarities in charge carrier processes (migration and diffusion) and the structure of the two oppositely charged regions in both electronic and iontronic diodes, we considered a similar model for the iontronic diode but with several modifications.

Figure 2(d) presents the measured current versus voltage (IV) characteristics of the iontronic diode obtained by quasi-steadily sweeping the applied voltage. A linear function can be fitted to describe the relationship between the current and voltage in two different operating regions: negative and positive voltage. This relationship can also be described by a voltage-dependent resistor. Figure 2(e) depicts the diode's total resistance under steady state condition $R_{TSS}$, extracted from the IV curve, as a function of the applied voltage $V_{in}$.

In the iontronic diode model, shown in Figure 2(b), the diode D1 from the electronic PN model, which has an exponential current-voltage relationship, is replaced by a voltage-dependent resistor $R_p(V)$. A voltage-dependent capacitor $C_p(V)$ is connected in parallel with $R_p(V)$. In addition, like $R_s$ in the PN model, $R_e$ in the iontronic model represents a parasitic resistance primarily associated with the resistance of the interconnecting microchannels.

EIS measurements, which are commonly used to examine electrochemical devices, were made on the bipolar iontronic diode shown in Figure 1(c) at five different DC bias voltages (-1V, -0.5V, 0V, 0.5V, 1V) over a frequency range of 0.1 Hz to 1 MHz. The work presented by J. H. Han et al. [37] served as the basis for fitting the impedance data to the simplified equivalent circuit shown in Figure 2(c), which comprises two Constant Phase Elements (CPEs), also known as fractional capacitors.

In the frequency domain, the impedance of a fractional capacitor is defined as:

[1] $Z_{CPE} = \frac{1}{Y_0(jw)^\alpha}$  $0 < \alpha < 1$,

where $Y_0$ is the CPE constant and $\alpha$ is the dispersion coefficient. In the time domain, the classical capacitor equation's time derivative is replaced by a fractional derivative of order $\alpha$ (where $0 < \alpha < 1$):

[2] $I(t) = Y_0 \frac{d^\alpha v(t)}{dt^\alpha}$,

where $\frac{d^\alpha v(t)}{dt^\alpha}$ is the fractional derivative of order $\alpha$ of the voltage v(t).

According to the Caputo definition of a fractional derivative, this equation implies that the current through a fractional capacitor is not only proportional to the rate of change of voltage but also depends on the historical voltage changes over time [39]. The dispersion coefficient $\alpha$ directly influences these history-dependent effects. In Figure 2(c), R0, R1 and Y01, $\alpha$01 are associated with the microchannel and electrode-solution interfaces (both anode and cathode), while R2 and Y02, $\alpha$02 are associated with the polyelectrolyte junction.

Figure 2(f) illustrates how the circuit components' values vary with applied DC voltage. For example, the resistance associated with the diode's junction ($R_2$) increased by two orders of magnitude when switching from forward to reverse bias, unlike $R_0$ and $R_1$, which remained relatively constant as the solution's ohmic resistance in the microchannel and redox processes at the electrode-solution interfaces remined unchanged. Similarly, Y02 changed by two orders of magnitude from forward to reverse bias, with its dispersion coefficient ($\alpha$02) varying with the bias voltage. By contrast, Y01 and its $\alpha$01 remained unchanged, with values around a few pF and 1, respectively. The impact of $C_1$ was primarily observed at high frequencies, suggesting that for lower frequencies (frequencies well below 800 kHz), a simpler equivalent circuit could describe the system (Figure S1 (b)). This simpler circuit, consisting of a parallel RC circuit with a voltage-dependent resistor and a voltage-dependent CPE capacitance in series with a constant resistance ($R_0 + R_1$), closely resembles the proposed iontronic compact model in Figure 2(b). In this model, $R_0 + R_1$ corresponds to $R_e$, and $R_2$ corresponds to $R_p$, with the key difference being the use of a fractional capacitor (Y02 and $\alpha$02) and not a voltage dependent regular capacitor (Cp).

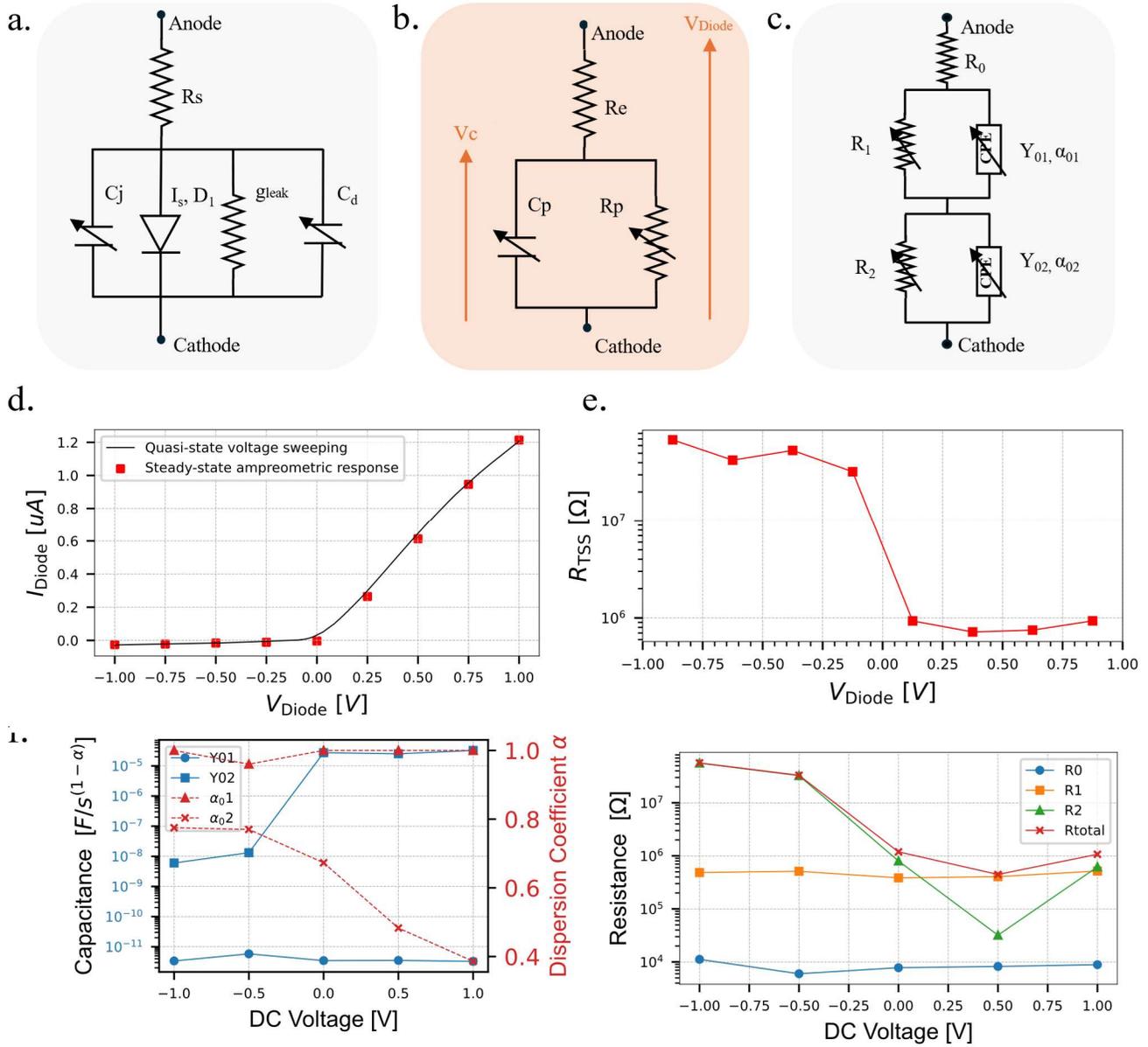

**Figure 2**: (a) **Electronic PN diode compact model based on the SPICE model** that served as the inspiration for the proposed iontronic diode model. (b) **The proposed iontronic diode compact model.** (c) **The equivalent circuit that was employed to analyze the electrochemical impedance spectroscopy (EIS) measurements.** (d) **Current-voltage (IV) characteristics of the iontronic diode**, determined by quasi-steady-state voltage sweeping (black line). Solid red squares indicate the steady-state response of the amperometric measurements under a constant applied V after 500 s. (e) **Total resistance ($R_{TSS}$) of the diode under steady-state conditions**, extracted from the IV curve, as a function of the applied voltage ($V_{Diode}$). (f) Variation of the EIS equivalent circuit component values with applied DC voltage ($V_{Diode}$).

## Phase II: Calibration

In simulations of any diode-based circuit, including logic gates, both the steady and transient response are crucial. We used the experimentally collected data shown in the I-V curve (Figure 2(d)) and the step response measurements (Figure 3(a)) of the diode to calibrate the proposed equivalent circuit parameters ($R_e$, $R_p(V)$, $C_p(V)$) (Figure 2(b)). While the I-V curve provides information on the quasi-steady state current, the step response measurement provides important information on transient behavior, in particular on the switching time of the diode.

Similar to PN semiconductor diodes that have two main operating regions, we defined the behavior of the voltage-dependent resistor and capacitor of the iontronic diode for two states: $Vc < 0$ and $Vc \geq 0$. (To more precisely capture the behavior of individual components, we modeled the voltage-dependent resistance and capacitance using $Vc$ instead of $V_{Diode}$. Further details are provided in the Method section). The experimental behavior and the interpretation of the suggested equivalent circuit reflected such separation of the volage dependency into two states as well.

$$[3] \quad R_p(V) \equiv \begin{cases} R_p^- & Vc < 0 \\ R_p^+ & Vc \geq 0 \end{cases}$$

$$[4] \quad C_p(V) \equiv \begin{cases} C_p^- & Vc < 0 \\ C_p^+ & Vc \geq 0 \end{cases}$$

To calibrate the five parameters required for the model ($R_e, R_p^-, R_p^+, C_p^-, C_p^+$), 5 constraints were defined. The resistor calibration stage involved setting 3 resistance constraints based on the steady state values of the obtained current under forward ($R_p^+$) and reverse bias ($R_p^-$), and the temporal overshoot current measured immediately after switching the voltage and representing the ohmic resistance ($R_e$) (more detailed information can be found in the Methods section).

To calibrate the capacitor $C_p$ we examined the current stabilization times for the different input voltage steps. The various voltage steps were designed to test the diode's transition between the different operation states (no bias 0V, forward bias 1V, and reverse bias -1V) (Figure 3(a)).

Based on the equivalent circuit model (Figure 2(b)), we inferred the data using a single exponential function corresponding to a single time scale. This was then used to write the two remaining constraints to calculate the capacitor values $C_p^+, C_p^-$ (more detailed information can be found in the Methods section).

After calibrating the parameters, we implemented the model in MATLAB to examine its suitability to assess the full dynamic electric behavior of the iontronic diode in various step voltage scenarios (Figure 3(a)). As expected, the model predicted all scenarios well, except for a single case involving the transition from forward to reverse bias. This transition exhibited more complex transient behavior that could not be fully captured by the simplified model. We speculate that this may be related to a memory effect [40], which is supported by the observation of fractional capacitance in the EIS measurements.

As one step towards the simulation of iontronic circuits containing several iontronic diodes, we successfully simulated the model in Cadence Spectre, a very common tool for simulating electronic circuits (Figure 3(a)), achieving results closely matching those obtained in MATLAB. For this purpose, the suggested voltage-dependent resistor and capacitor models were written in VerilogA, the de-facto language for defining and distributing compact models for electronic components [41].

## Model of process variation effects as the first add-on to the core model

Similar to electronic devices, iontronic diodes exhibit random variations in their physical characteristics related to the fabrication process. For instance, Figure 3(b) illustrates the variation in junction length (i.e., the polyelectrolytes contructing the iontronic bipolar diode) in 9 different diodes on the same chip. To accurately simulate an iontronic circuit, the diode model must account for these variations, since they influence the electrical behavior of each individual diode and, consequently, the overall circuit performance significantly.

In this work, we propose using a Monte Carlo (MC) simulation technique [42] to evaluate the functionality of integrated iontronic circuits, and specifically as regards intra-chip variations. Inter-chip variations which involve global factors that impact all devices identically on a chip (such as the solution, polymer, or material properties), are beyond the scope of this study, since assessing them would require fabricating and measuring multiple chips. By conducting numerous random simulations, the Monte Carlo technique estimates circuit performance and yields in a range of possible conditions, thus allowing designers to make informed decisions that lead to more reliable and manufacturable designs.

Major variations were observed in the I-V curves (Figure 3(c)), particularly in the distribution of quasi steady-state currents under reverse bias ($I_{off}$ at -1V) and forward bias ($I_{on}$ at 1V) for each diode. A log-normal distribution was fitted to describe the results for both $I_{on}$ and $I_{off}$. This behavior was also observed in another iontronic integrated chip with a larger number of diodes (Figure S2).

We attribute these variations primarily to inconsistencies in the diode junction dimensions, in particular related to $R_p$. Given that the solution, electrodes, and microchannel geometry can be controlled with precision, we assumed a constant $R_e$. Consequently, the statistical distributions of $R_p^+$ and $R_p^-$ were derived from the measured steady-state currents ($I_{on}$ and $I_{off}$) across 15 diodes. For instance, the statistical distribution of Rp+ is shown in Figure 3(d). In our work, to analyze an iontronic circuit, we perform hundreds of simulation runs. For each MC run, the values of $R_p^+$ and $R_p^-$ for each diode in the circuit are randomly selected according to the specified distribution. The circuit's functionality and performance are then examined as a function of this variation. We note that this approach, which allows for the execution of numerous simulations to explore parameter variations, is advantageous compared to tools like COMSOL, where running hundreds of simulations to study such variations is not feasible. This highlights the necessity of using compact models for efficient analysis.

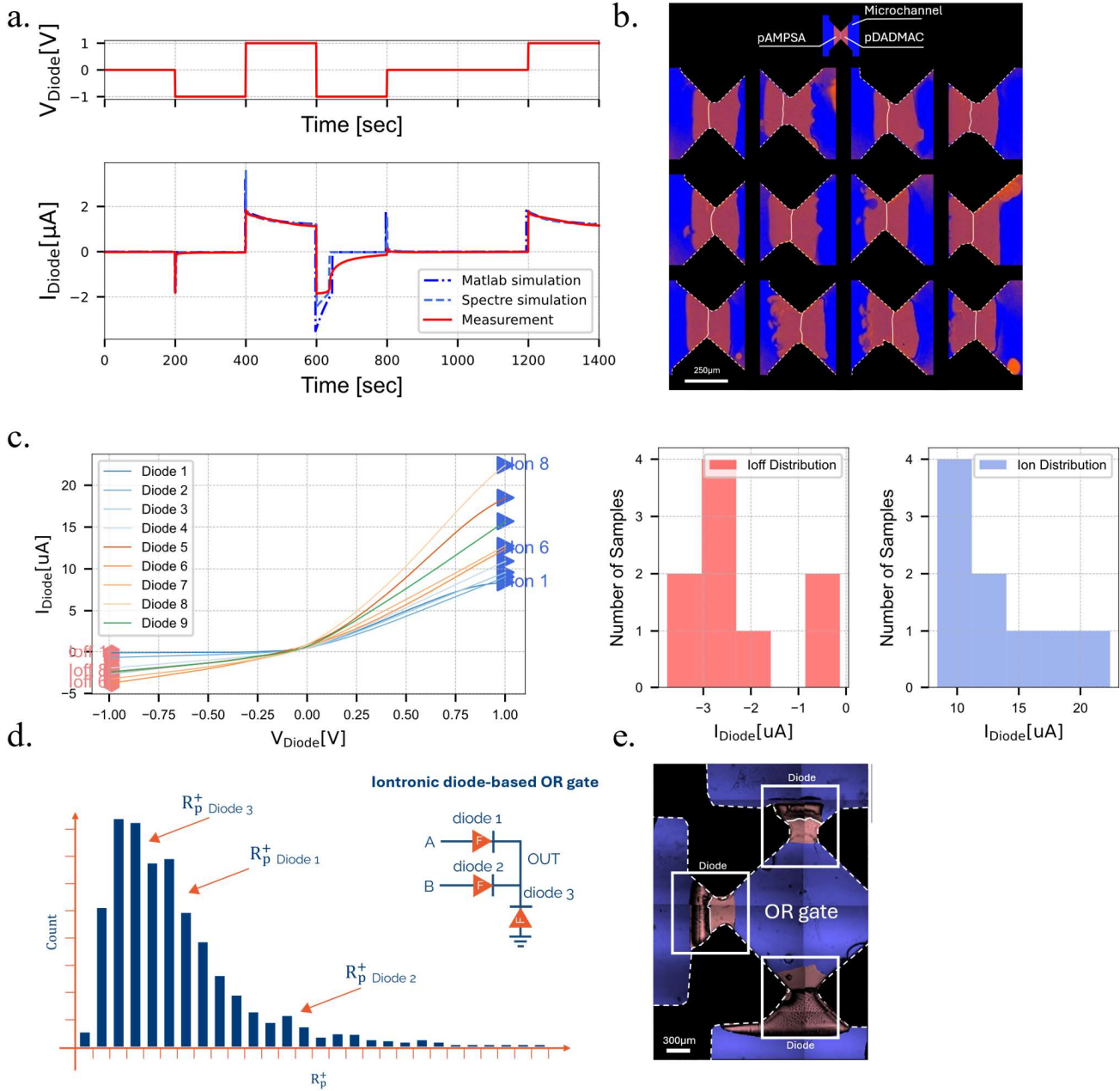

**Figure 3:** (a) **Step-response measurement and simulation**: The behavior of the iontronic diode's current ($I_{Diode}$) as a function of time for 4 different transitions of the input voltage ($V_{Diode} = 0, -1, 1\ V$). The red line represents the experimental results, and the blues lines represent the simulation results using two different tools – Matlab and Spectre. (b) **An image of 9 fabricated diodes:** the variation in the physical parameters in the polyelectrolyte junction area (marked in red), can be seen. The microchannels with the solution are marked in blue (c) **I-V variations:** the variations observed in the electrical characteristics of the 9 diodes. The differences were caused by variations during chip fabrication. $I_{off}$ and $I_{on}$ are the measured diode's current when applying a $V_{diode}$ equal to -1V and 1V, respectively. On the right, the distribution of $I_{off}$ and $I_{on}$ can be seen. (d) **The basic concept of the Monte Carlo iontronic circuit simulation:** For each diode in the circuit, a random value was generated to Rp+ according to a specified statistical distribution. The figure also presents an example of arbitrary Rp+ values chosen for 3 different diodes constructing an OR gate in a single MC run. **(e)** Image of the fabricated iontronic OR gate consisting of 3 iontronic diodes (marked in red) and interconnected microchannels (marked in blue).

## Circuit simulation with iontronic diodes

Designing an electrical circuit involves shifting the focus from individual components to their interactions. Instead of viewing components as isolated, they are evaluated based on the ways they affect the overall circuit. As circuits get more complex, predicting these interactions and their impact on performance becomes more difficult. The design process centers on achieving the desired circuit output by managing these interactions.

A Spectre Cadence circuit simulation tool utilizing our diode model, was used to test the feasibility of constructing logic gates solely with iontronic diodes with nominally identical properties. Figure 4(a) illustrates the implementation of AND and OR logic gates and compares the measurement results taken from Sabbagh et al. [28] with the Monte Carlo simulation outcomes.

Expanding this case study to a larger integrated circuit consisting of five OR gates (15 diodes in total) connected in series allowed us to evaluate the simulation's accuracy in predicting the behavior of more complex circuits with interconnected logic gates. As shown in Figure 4(b), despite the considerable difficulties in predicting the performance of such a cascade of diode-based logic gates (The fabricated chip is shown in Figure 4(c)), the experimental results were within three standard deviations ($3\sigma$) of the simulated data, indicating good agreement between the observed and predicted values.

To assess the potential for cascading circuits based on the iontronic diode, we simulated the maximum number of OR gates that could be cascaded. Success was defined as an output voltage of the last gate exceeding 500 millivolts, which thus served as the threshold between different logic levels. To ensure a high production yield, we required a 99% success probability in the simulations. We examined different logic chain lengths of up to 34 gates and the influence of various diode parameters ($R_e$, $R_p^-$, $R_p^+$) on the cascading ability. These parameters affected the diode's Rectification Ratio (RR), which is defined as the ratio of steady-state resistance in reverse and forward bias.

Figure 4(d) shows that increasing the RR allowed for more cascaded gates. Reducing the component variability while increasing the RR resulted in greater improvement. This is shown by the orange line in Figure 4(d), representing the simulation results, with a standard deviation of $R_p^-$, $R_p^+$ ten times smaller than measured on the fabricated chip. For example, designing an iontronic logic chain with 15 gates, while achieving a standard deviation reduced by an order of magnitude, would require increasing the RR by only 25 instead of 100. This highlights the critical importance of component uniformity in more complex circuits.

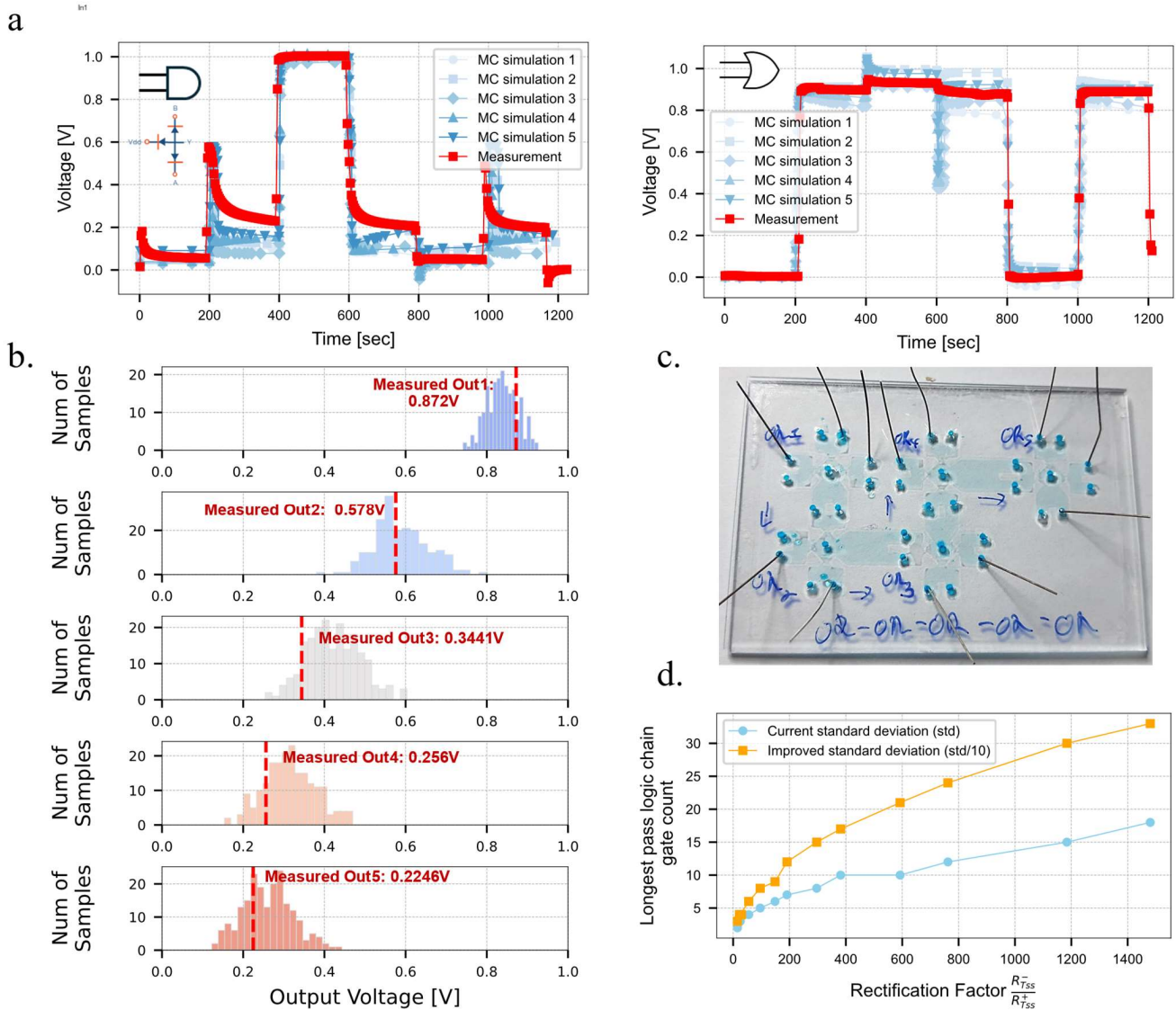

**Figure 4:** (a) **Logic Gate Simulation and Measurement:** Measurements (taken from Sabbagh et al. [28]) and 5-point Monte Carlo simulations of AND and OR gates using the iontronic compact model. (b) **Cascaded OR Gates:** A 200-point Monte Carlo simulation of five cascaded OR gates, where each simulation point produced a different circuit output due to statistical variations. The resulting output voltages for each gate in the chain are summarized as distributions. The top plot shows the distribution for the first gate, and the bottom plot for the last gate in the chain. The measurement results are represented as dashed lines overlaid on these distributions, allowing for a direct comparison between the measured and simulated data. (c) **An image of the fabricated iontronic chip with 5 integrated OR gates** (d) **Logic Chain Length vs. Rectification Ratio:** Analysis of the maximum operational logic chain length as a function of the Rectification ratio. The x-axis represents the Rectification Ratio, as the ratio between the reverse to the forward steady-state resistance, and the y-axis shows the longest operational logic chain identified. The blue line indicates the standard deviation as measured on the fabricated chip, while the orange line shows the case where variations are reduced; here, the standard deviation of $R_p^+$ and $R_p^-$ was 10 times smaller (std/10).

Building significant logic modules require implementing an inverter (NOT) gate. Since this is not feasible with diodes alone, we used Dual Rail Logic, and represented each bit with two logic signals [43]. This method doubles the required gates but makes it possible to design complex modules without an inverter gate. We designed an iontronic decoder based on two-input dual-rail AND gates (overall the circuit was composed of 24 gates).

A decoder, which is mandatory in electronic systems, converts binary code into a unique representation of values. An iontronic decoder could potentially control an ion-based memory array or be used as a control circuit for different innovative sensor arrays on an iontronic chip.

Figure 5(b) shows a simulation of a 3x8 (3-inputs, 8-outputs) iontronic decoder (Figure 5(a)), that only activates the corresponding output upon receiving binary input. We examined through the simulations the margin between the high output to the highest low output (High-to-Low margin), the time period it took the output signal to settle after input change (settle time), and the decoder's power consumption. Figure 5(c) shows that reverse resistance ($R_p^-$,) significantly affected power consumption and the High-to-Low margin. Increasing $R_p^-$, greatly improved output logic, and enhanced the High-to-Low margin. A tenfold increase in $R_p^-$, improved all the decoder performance parameters and had greater impact than decreasing the forward resistance ($R_p^+$).

Implementing a fully iontronic circuit that functions independently of traditional electronic components would likely be a groundbreaking step forward [16]. This approach opens up new pathways for energy-efficient, biocompatible, and potentially self-sustaining systems that leverage ion-based principles without the constraints of conventional electronics. One theoretical solution is an iontronic circuit designed for harvesting radio frequency (RF) energy. The key feature of this system is a diode bridge (rectifier bridge), which ensures that the current flows in only one direction (Figure 5(d)), and allows both halves of the AC waveform to be used, thus producing a pulsating DC output that can be then converted to a direct current using a capacitor. In [44] A possible implementation of such ionic rectifier based on nanochannels was presented by J. Li et al. [44].

Currently, the response times of the existing iontronic diodes are excessively long, and do not meet typical RF frequency requirements. However, the developed simulations enabled us to explore the maximum operating frequency and assess how changes in the diode's resistance or capacitance can impact its performance.

In the current diode characterization, the maximum operating frequency is 0.1mHz. Therefore, as shown in Figure 5(e), the output signal (with the $C_p$ multiplication equal to 1) fails to follow the input signal (with an input signal frequency of 0.1Hz). However, reducing the diode's capacitance ($C_p$) by a factor of 1,000 can increase the maximum operating frequency to 0.1 Hz and significantly enhance the circuit's functionality.

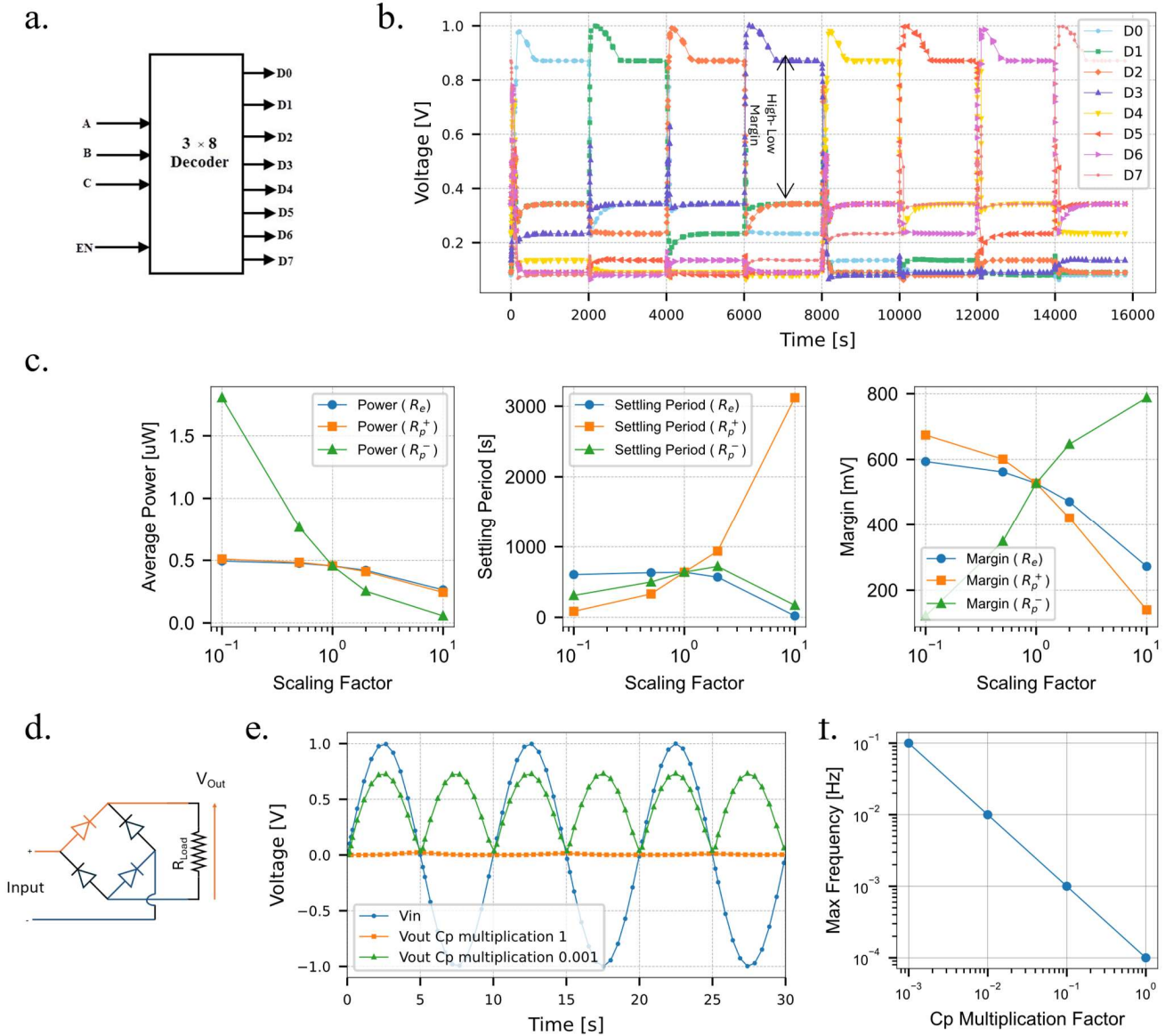

**Figure 5:** (a) **3-to-8 Decoder:** Schematic of a 3-to-8 decoder that converts a 3-bit binary input (A,B,C) into a single active output from eight possible outputs (D0-7). The iontronic decoder is designed with 2-input dual rail AND gates and includes overall 24 iontronic logic gates. (b) **Iontronic Decoder Simulation**: Waveforms showing the High-Low Margin; i.e.,the voltage difference between the selected output (high) and the non-selected outputs (low). This margin indicates the decoder's performance in distinguishing between active and inactive outputs. (c) **Quality Metrics:** Plots of various performance metrics of the iontronic decoder: (from left to right) average power consumption during operation (including both switching and static periods), settling time of the output signal after an input change, and the High-Low Margin. All were assessed across different decoder parameters. (d) **Diode Bridge Circuit:** Circuit diagram of a diode bridge connected to a resistor, which converts the AC input into a pulsating DC output. This pulsating DC can be smoothed into a direct current with the addition of a capacitor. (e) **Iontronic Diode Bridge Simulation based on (d) for varying Cp values (orange and green lines) under an oscillating input voltage at 0.1Hz (blue line)** (f) **Frequency Response:** Plot showing the maximum operational frequency of the iontronic diode as a function of the $C_P$ multiplication factor, where $C_{P\_new}$ = multiplication factor × $C_P$.

## Conclusions

This paper presented a design approach toward future large-scale iontronic integrated circuits, inspired by electronic design methodologies. As a foundational step, we introduced a compact model for an iontronic diode. Although demonstrated here with iontronic bipolar diodes, this method is broadly applicable and can extend to other iontronic components. The model, based on the diode's physical properties and observed behavior, was implemented using Verilog-A within the Spectre circuit simulator, allowing integration with standard VLSI tools.

The simulation results provide insights into the behavior of diode-based iontronic circuits, and emphasize how the parameters of individual components affect circuit interactions and overall performance. Incorporating a Monte Carlo technique from electronic circuit design further improved the model's accuracy in accounting for variations within iontronic circuits.

The expansion of the circuit simulation tool to include iontronic component provides valuable insights into the design and potential of iontronic circuits and helps explore their potential applicability in energy harvesting and advanced logic circuits. By improving component uniformity, optimizing diode parameters and better understanding diode characteristics, we can achieve more complex and efficient iontronic systems, thus moving closer to practical and groundbreaking applications in the field.

Developing a physics-based model that incorporates physical parameters such as the iontronic component's physical dimensions, the membrane's ion-permselectivity, the solution ion concentrations, etc. is likely to lead to significant advantages in terms of reliability, scalability, and predictive capability. The electrochemical impedance spectroscopy (EIS) measurements as well as the current-voltage (I-V) scans conducted in this study represent an important step toward this goal, by offering valuable insights into the processes within the diode. One key finding emerging from these measurements is the need to integrate a fractional element in future models to fully capture the diode's behavior, as well as accounting for parasitic resistances associated with the interconnecting microchannels. While there are various approaches to incorporating this fractional element into electronic circuit simulators, further research is needed to determine the most effective method, especially for a fractional capacitor that can vary with the applied voltage. Our future efforts will concentrate on this area to enhance both our understanding and the accuracy of the model.

Another critical area for future work involves developing simulation tools that address the dynamics of the current caused by the movement of multiple ion types in iontronic components. Since different ions can simultaneously contribute to the electric current—each potentially moving at its own velocity—new, complex behaviors may emerge in iontronic circuits that are not yet fully explored in current simulations or applications. Overcoming the challenges of designing circuits that leverage this complexity will be essential for advancing iontronic technology.

## Acknowledgements

This work was supported by the Israel Innovation Authority (IIA) and the Israel Science Foundation (Grant No. 1934/20). The authors express their gratitude to Prof. Danilo Demarchi for the discussions that inspired the direction of this research. The authors also thank Dr. Yuval Edri for the valuable discussions on calibrating the model parameters and for his contribution in developing the MATLAB framework.

## Methods

For more detailed information on the diode structure, measurements and fabrication process, the reader is referred to our previous work [28].

### Calibration of the model parameters

To calibrate the five parameters required for the model ($R_e, R_p^-, R_p^+, C_p^-, C_p^+$), 5 constraints were written.

**Resistor calibration**

The first two constraints were written by observing the I-V measurements (Figure 2(d)). We assumed that in the steady state condition, the capacitor $C_p(V)$ was in a state of very high resistance (cutoff) so that the entire diode current would only flow through the resistors $R_e$ and $R_p(V)$. In this case, the total diode resistor in the steady state, $R_{Tss}(V)$ was equal to the connection of the two resistors. The value of $R_{Tss}(V)$ was   evaluated from the slope of the I-V characteristics.  The deducible resistance value as a function of the voltage can be seen in Figure 2 (e). Recall that there are two main operating regions: when the diode voltage $V_{in} = V_{Diode}$ is greater than zero (forward bias) and when $V_{Diode}$ is less than zero (reverse bias). With this, there are changes in $R_{Tss}(V)$ within the operating regions, which however are relatively small compared to the changes in the values of the resistance in forward and reverse bias. The mean values in both bias conditions were calculated and used to write the first two constraints:

[1] $R_{Tss}^- = R_e + R_p^- = 4.9 \cdot 10^7 \Omega$   $V_{Diode} < 0$

[2] $R_{Tss}^+ = R_e + R_p^+ = 8.4 \cdot 10^5 \Omega$   $V_{Diode} \geq 0$

For simplicity we took the value at 0 to be equal to the total resistance for the forward bias. $Vc$, the voltage across $R_p(V)$ and $C_p(V)$, was defined as the diode voltage ($V_{Diode}$) minus the voltage drop across $R_e$, which was assumed to be constant in our model. When $V_{diode}$ exceeded zero, $Vc$ also became positive. To more accurately represent the behavior of individual components, we modelled the voltage-dependent $R_p(V)$ and $C_p(V)$ based on $Vc$ rather than $V_{Diode}$, and preserving the regional separation for positive and negative values. To find the parameters of $R_p(V)$, we first calibrated $R_e$ by using the current measurement as a function of time for the various voltage steps. In the diode measurement, overshoot where the current exceeds the final value during the transition response, was observed. For the electronic circuit, which was used as the model for this diode, the overshoot was due to the presence of the capacitor $C_p(V)$. We assumed that at the moment of change, the capacitor was short, and that the total resistance of the circuit was equal to $R_e$. We extracted the highest current value, $I_{peak}$, after changing the diode voltage from zero to 1V, to produce the third constraint:

[3] $R_e = \frac{V_{in}}{I_{peak}} = 5.5 \cdot 10^5 \ \Omega$

This constraint served to calculate $R_p^-$ and $R_p^+$ from the first two equations.

[4] $R_p^+ = 2.9 \cdot 10^5 \ \ \Omega$

[5] $R_p^- = 4.84 \cdot 10^7 \ \Omega$

**Calibration of the voltage dependent capacitor parameters**

Each transition was measured twice to ensure repeatability.

Given the electrical circuit described in Figure 2(b) the system can be described using the following single ordinary differential equation:

[6] $\frac{dQ}{dt} = \frac{V_{in}}{R_e} - \frac{Q}{C_p} \cdot \frac{R_e + R_p}{R_p R_e}$

where Q represents the charge on the capacitance $C_p$, and $V_{in}$ is the diode voltage.

The analytical solution of the above differential equation is:

[7] $Q = \frac{V_{in} C_p R_p}{R_e + R_p} + A \cdot e^{-\left(\frac{R_p + R_e}{C_p R_p R_e}\right)t}$

where A is a constant which is determined by the initial condition. From equation [7] and the full relationship between Q and the current on $R_e$:

[8] $I_{R_e} = I_C + I_{R_p} = \frac{dQ}{dt} + \frac{V_C}{R_p} = \frac{dQ}{dt} + \frac{Q}{C_p R_p}$

We get the general solution for the diode current (which is equal to the current on $R_e$):

[9] $I_{diode} = I_{R_e} = \frac{V_{in}}{R_e + R_p} - \frac{A}{C R_e} \cdot e^{-\left(\frac{R_p + R_e}{C_p R_p R_e}\right)t}$

This predicts a growth / decay curve with a single time scale:

[10] $\tau = \frac{C_p R_p R_e}{R_p + R_e}$

Based on the model, we inferred the data using a single exponent. The current responses for the different step scenarios are plotted in **Error! Reference source not found.**(a). After normalizing by subtracting the steady state value from the signal measurement and then dividing it by the value of the current minus the steady state value at $t = 0$ we get:

[11]
$$I_{diode_{manipualte}} = \frac{I_{diode} - I_{ss}}{I_{diode}(t=0) - I_{ss}} = e^{-\left(\frac{R_p + R_e}{C_p R_p R_e}\right)t} = e^{-\frac{t}{\tau}}$$

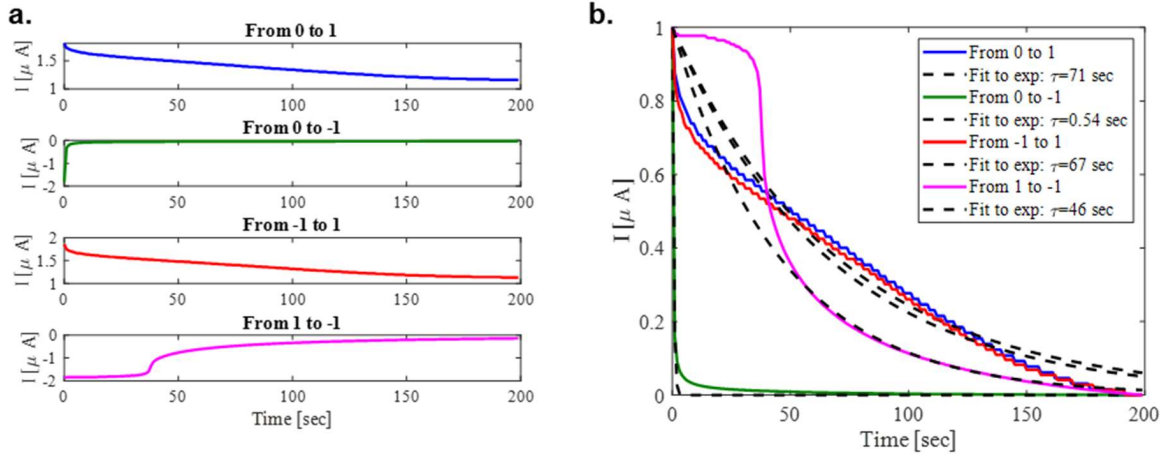

**Figure M 1**: (a) Step-response measurements of the iontronic diode's current as a function of time for four different input voltage transitions presented above (Figure 3(a)). In this figure, the original measurement was divided into four separate graphs, each isolating and highlighting the detailed behavior of an individual transition. (b) The normalized diode current is displayed as solid curves for each transition, with the corresponding exponential fits shown as dotted lines.

The result of the mathematical manipulation of the diode current measurement is depicted in Figure M1(b) as solid curves (each curve corresponds to a different input voltage transition). Each dotted curve represents the exponential fit to the $I_{diode_{manipualte}}$ of the same transition.

We obtained: $\tau \approx 70\ sec$ for the steps between 0 to 1 V and for steps between -1 to 1 V, $\tau \approx 0.54\ sec$ for steps between 0 to -1 V, and $\tau \approx 46\ sec$ between -1 V to 1 V.

This yielded two more equations:

[12]        $\tau^{0,1} = \frac{C^+ R_p^+ R_e}{R_p^+ + R_e} = 71\ sec$

[13]        $\tau^{0,-1} = \frac{C^- R_p^- R_e}{R_p^- + R_e} = 0.54\ sec$

Note that the timescale for the decay curve for a response from 0 to -1 was much shorter than the response to steps between 1 to -1. We assumed that this was the result of memory such that when moving from 1 to -1 the capacitor stayed at the $C^+$ value, which should then follow the value:

[14]        $\tau^{1,-1} = \frac{C^+ R_p^- R_e}{R_p^- + R_e}$

However since $R_p^+ < R_p^-$, according to Eq [12] and Eq [14], $\tau^{0,1} < \tau^{1,-1}$, did not match the experimental value. Thus, to capture this experimental feature, more complexity needed to be added to the model.

Given the calibrated values for the resistors, the value of the capacitor was calculated as:

$$[15] \qquad C^+ = 3.74 \cdot 10^{-4} [F]$$

$$[16] \qquad C^- = 9.93 \cdot 10^{-7} [F]$$

## large language model

We acknowledge the use of a large language model (LLM) for grammar and improving the clarity of the language in this manuscript. Additionally, the LLM was employed to assist in developing scripts for visualizing simulation and experimental data.